\documentclass[sigconf]{acmart}

\usepackage{multirow}

\usepackage{enumitem}

\usepackage{subcaption}

\usepackage{float}  

\usepackage[table]{xcolor}

\AtBeginDocument{%
  }

\setcopyright{none}
\acmDOI{}
\acmISBN{}

\begin{document}

\title{Beyond One-Size-Fits-All: Adaptive Test-Time Augmentation for Sequential Recommendation}

\author{Xibo Li}
\affiliation{%
  \institution{The Hong Kong University of Science and Technology (Guangzhou)}
  \city{Guangzhou}
  \country{China}
}
\email{lixibo0707@gmail.com}

\author{Liang Zhang}
\authornote{Corresponding author.}
\affiliation{%
  \institution{The Hong Kong University of Science and Technology (Guangzhou)}
  \city{Guangzhou}
  \country{China}
}
\email{liangzhang@hkust-gz.edu.cn}


\renewcommand{\shortauthors}{Anonymous Submission}


\begin{abstract}
Test-time augmentation (TTA) has become a promising approach for mitigating data sparsity in sequential recommendation by improving inference accuracy without requiring costly model retraining. However, existing TTA methods typically rely on uniform, user-agnostic augmentation strategies. We show that this ``one-size-fits-all'' design is inherently suboptimal, as it neglects substantial behavioral heterogeneity across users, and empirically demonstrate that the optimal augmentation operators vary significantly across user sequences with different characteristics for the first time. To address this limitation, we propose AdaTTA, a plug-and-play reinforcement learning–based adaptive inference framework that learns to select sequence-specific augmentation operators on a per-sequence basis. 
We formulate augmentation selection as a Markov Decision Process and introduce an Actor–Critic policy network with hybrid state representations and a joint macro–rank reward design to dynamically determine the optimal operator for each input user sequence.
Extensive experiments on four real-world datasets and two recommendation backbones demonstrate that AdaTTA consistently outperforms the best fixed-strategy baselines, achieving up to \textbf{26.31\%} relative improvement on the Home dataset while incurring only moderate computational overhead ($\sim\!1.48\times$).

\end{abstract}



\maketitle

\section{Introduction}

Sequential recommendation (SR) has attracted substantial attention due to its close alignment with real-world user interaction scenarios \cite{kang2018self, zhai2025combinatorial}. Over the past decades, existing SR models have demonstrated strong capability in capturing transition patterns and user preferences from historical interaction sequences \cite{hidasi2015session, tang2018personalized}. However, their effectiveness is fundamentally limited by the severe data sparsity issue inherent in user behavior, as most users interact with only a small number of items \cite{dang2024data, meng2023parallel}. To alleviate this challenge, researchers have proposed various data augmentation methods, including heuristic sequence expansion \cite{tang2018personalized}, model-based augmentation methods such as counterfactual generation and diffusion models \cite{wang2021counterfactual, liu2023diffusion, liu2021augmenting}, as well as contrastive transformations \cite{xie2022contrastive, liu2021contrastive}.

Nevertheless, most existing augmentation methods predominantly operate during the model training stage. While effective, deploying these approaches typically requires retraining models from scratch or fine-tuning them with augmented data, which can be prohibitively time-consuming and computationally expensive—especially for large-scale datasets and modern recommendation models \cite{dang2025data}. Test-Time Augmentation (TTA) offers a promising alternative to address these limitations \cite{shanmugam2020and, zhang2022memo}. Instead of modifying the training process, TTA augments input sequences during inference and aggregates predictions across augmented variants to improve recommendation performance. Importantly, TTA introduces no changes to model architecture and requires no retraining, making it highly attractive for real-world deployment. Recently, Dang et al. \cite{dang2025data} pioneers the application of TTA to sequential recommendation, proposing operators such as TNoise and TMask that achieve performance comparable to retraining-based methods. 

However, as illustrated in Figure~\ref{fig:introduction_figure} (left), existing TTA approaches adopt a fixed and user-agnostic augmentation strategy, applying the same operator uniformly across all test sequences. We argue that such a one-size-fits-all design is inherently suboptimal. User interaction sequences exhibit substantial heterogeneity, and different sequences may benefit from different augmentation operations (\textit{e.g.}, crop, mask or substitute). Consequently, indiscriminately applying a single operator fails to fully exploit the potential of TTA. Therefore, in this work, we investigate \emph{adaptive test-time augmentation} for sequential recommendation for the first time and seek to answer two fundamental research questions:
\begin{itemize}[leftmargin=*,noitemsep]
\item \textbf{Q1:} Does a uniform augmentation strategy remain optimal across heterogeneous user sequences?
\item \textbf{Q2:} Can a learning-based framework automatically select the optimal augmentation operator conditioned on the characteristics of an input sequence?
\end{itemize}

\begin{figure}[htbp]
    \centering
    \includegraphics[width=0.99\columnwidth]{Intro_2.pdf}
    \caption{Comparison between one-size-fits-all and sequence-adaptive test-time augmentation strategies. (Left) Existing TTA methods apply a fixed augmentation operator uniformly across all user sequences. (Right) Our proposed AdaTTA framework adaptively selects operators based on sequence characteristics.}
    \label{fig:introduction_figure}
\end{figure}

To answer \textbf{Q1}, we first conduct an empirical study to systematically analyze the performance of various augmentation operators on sequences with different characteristics. Our results reveal that no single operator is universally optimal; instead, the most effective augmentation strategy varies significantly with sequence properties such as length and interaction patterns. These findings empirically validate the necessity of sequence-adaptive augmentation.

Motivated by this observation and to answer \textbf{Q2}, we propose \textbf{AdaTTA}, a reinforcement learning–based framework for adaptive test-time augmentation, as illustrated in Figure~\ref{fig:introduction_figure} (right). Specifically, AdaTTA formulates augmentation operator selection as a Markov Decision Process (MDP), where a policy network observes a state representation derived from the input sequence and selects an augmentation action, with the goal of maximizing the expected improvement in recommendation performance through a joint macro–rank reward design. We conduct extensive experiments on multiple public benchmarks and a diverse set of representative sequential recommendation models. The results demonstrate that AdaTTA consistently outperforms all fixed TTA strategies. 
In summary, our main contributions are threefold:


\begin{itemize}[leftmargin=*,noitemsep]
    \item Conceptually, we systematically investigate adaptive test-time augmentation for sequential recommendation for the first time and reveal the inherent limitations of traditional one-size-fits-all TTA strategies.
    \item Methodologically, we propose AdaTTA, a reinforcement learning–based framework with a joint macro–rank reward design that enables sequence-adaptive augmentation operator selection.
    \item Empirically, we conduct extensive experiments across multiple public datasets and representative sequential recommendation models, demonstrating the effectiveness, robustness, and generalizability of AdaTTA framework.
\end{itemize}

\section{RELATED WORK}
\label{sec:related_work}

\subsection{Sequential Recommendation}

Sequential recommendation (SR) is an important task, aiming to predict the next item to access based on a sequence of interacted items~\cite{zhao2023sequential, zhao2023cross}. Early works leveraged Markov Chains~\cite{he2016fusing, rendle2010factorizing} to model user behaviors, where the next item to predict is closely related to the latest few interactions. Later, researchers adopted CNNs~\cite{yuan2019simple} and RNNs~\cite{hidasi2015session} to capture the relationships among items. For example, GRU4Rec~\cite{hidasi2015session} trained a Gated Recurrent Unit (GRU) architecture to model the evolution of user interests. Caser~\cite{tang2018personalized} embeds a sequence of recent items in the time and latent spaces, then learns sequential patterns using convolutional filters. Due to the success of the self-attention mechanism~\cite{vaswani2017attention}, a series of Transformer-based SR models have been proposed. SASRec~\cite{kang2018self} applied the Transformer layer to learn item importance in sequences, which characterize complex item transition correlations. BERT4Rec~\cite{sun2019bert4rec} further advanced this direction by employing bidirectional self-attention with masked item prediction. STOSA~\cite{fan2022sequential} embeds each item as a stochastic Gaussian distribution and devises a Wasserstein Self-Attention module to characterize item-item relationships. Beyond Transformer-based models, researchers found that filtering algorithms can alleviate the influence of the noise in SR models~\cite{zhou2022filter}. They proposed an all-MLP architecture with learnable filters. While these approaches focus on improving model architectures or training procedures, our work is orthogonal in that it enhances sequential recommendation at inference time through adaptive test-time augmentation, without modifying the underlying model.


\subsection{Data Augmentation for Recommendation}

Data augmentation has been shown to be an effective way of mitigating the data sparsity problem in SR~\cite{dang2024repeated, dang2025augmenting}. Early work adopted heuristic methods, such as Sliding Windows~\cite{tang2018personalized} and random replacing~\cite{wang2021counterfactual}, which split one sequence into many sub-sequences or augment based on counterfactual thinking. Later, some work explored using self-supervised signals to uncover richer preference information. For example, CL4SRec~\cite{xie2022contrastive} constructed contrastive views by three operators, \textit{i.e.}, Crop, Mask, and Reorder. CoSeRec~\cite{liu2021contrastive} further improved CL4SRec by presenting two informative operators, Substitute and Insert. In addition to modifying sequences by operators, some work utilizes generative models or specialized modules for augmentation. In this line, DiffASR~\cite{liu2023diffusion} adopted the diffusion model for sequence generation. Two guide strategies are designed to control the model to generate the items corresponding to the raw data. ASReP~\cite{liu2021augmenting} employed a reversely pre-trained transformer to generate pseudo-prior items for short sequences. 
However, deploying existing augmentation methods typically requires retraining models, modifying architectures, or introducing additional parameters, which limits their practicality. Recent work has investigated test-time augmentation (TTA) as an alternative paradigm. Dang et al.~\cite{dang2025data} systematically explored applying augmentation operators during inference stage rather than training phase, revealing that mask and substitute operators achieve superior performance by introducing appropriate perturbations. Despite these promising results, existing TTA approaches adopt fixed augmentation strategies that apply the same operator uniformly across all input sequences, leaving the potential of sequence-adaptive test-time augmentation largely unexplored.

\subsection{Reinforcement Learning in Recommendation}

Reinforcement learning (RL) reformulates recommendation as sequential decision-making problem to optimize long-term user engagement rather than immediate rewards~\cite{chen2019top}. Early work pioneered the application of deep reinforcement learning to news recommendation, demonstrating effective handling of dynamic content and user interests through Deep Q-Networks. Ren et al.~\cite{ren2023contrastive} advanced this by incorporating contrastive state augmentations for RL-based recommender systems, showing that augmented state representations provide valuable signals for policy learning. Chen et al.~\cite{chen2019top} addressed the challenges of applying REINFORCE to recommendation systems by proposing off-policy correction methods that enable learning from logged data. In contrast to prior RL-based recommender systems that directly learn recommendation policies, our work leverages reinforcement learning to select test-time augmentation operators, serving as an inference-time enhancement module.


\section{PRELIMINARIES}
\subsection{Sequential Recommendation}
\label{sec:problem}

Given a user set $\mathcal{U}$ and an item set $\mathcal{V}$, each user $u \in \mathcal{U}$ is associated with a sequence of interacted items in chronological order $s_u = [v_1, \ldots, v_j, \ldots, v_{|s_u|}]$, where $v_j \in \mathcal{V}$ indicates the item that user $u$ has interacted with at time step $j$ and $|s_u|$ is the sequence length. Then, sequential recommendation aims to accurately predict the most likely item $v^*$ that user $u$ will interact with at next time step $|s_u|+1$, formulated as:
\begin{equation}
    \arg\max_{v^* \in \mathcal{V}} P(v_{|s_u|+1} = v^* \mid s_u),
    \label{eq:sr_objective}
\end{equation}
where the item with the highest predicted probability will be selected for recommendation.

\subsection{Sequence Data Augmentation Operators}
\label{sec:operators}

\begin{figure}[htbp]
    \centering
    \includegraphics[width=0.99\columnwidth]{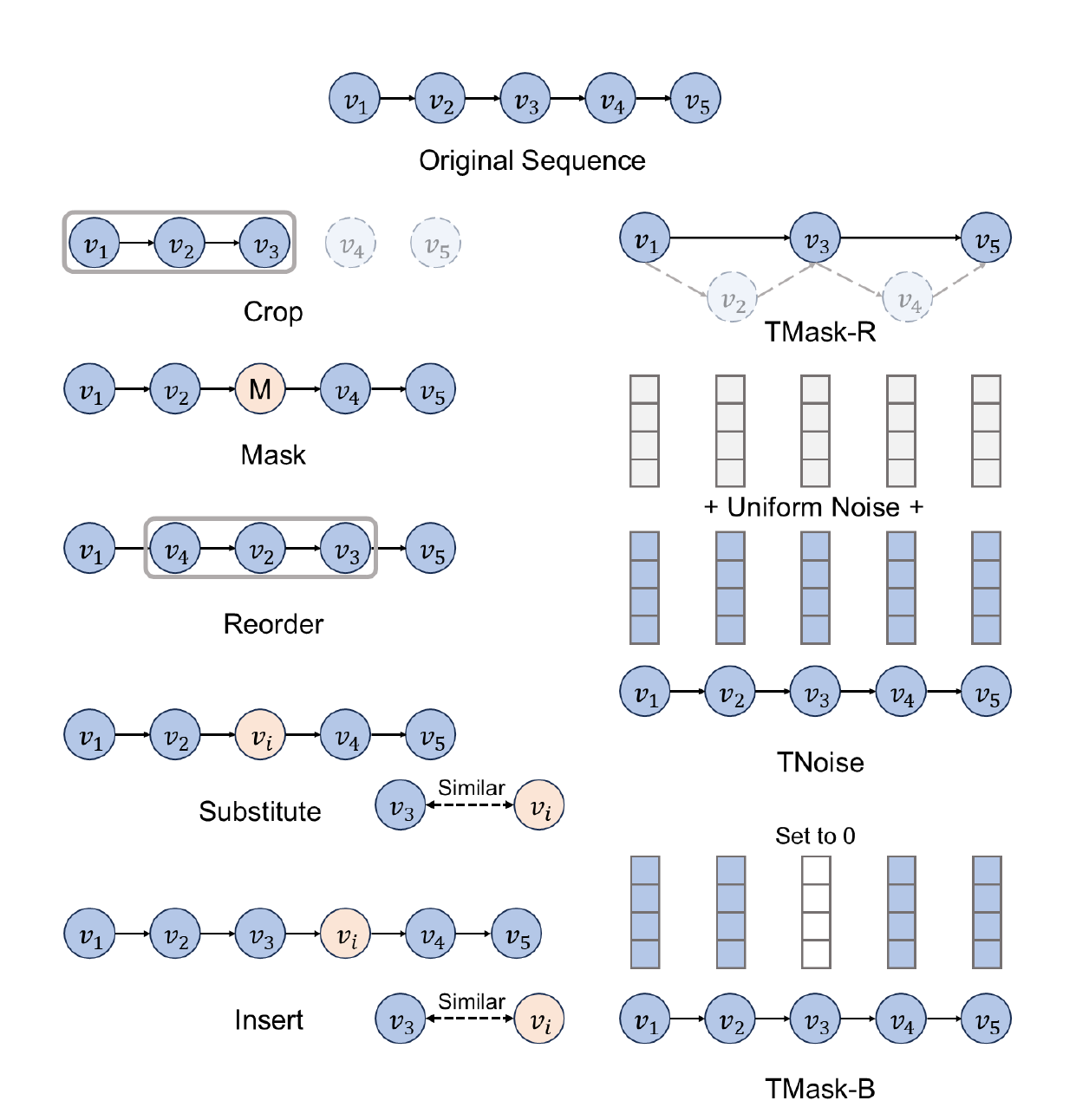}
    \caption{Illustration of data augmentation operators for sequential recommendation. The original sequence undergoes various transformations (e.g., masking, cropping, substitution, reordering, noise addition) to generate augmented views for improving model robustness and performance.}
    \Description{Illustration showing different data augmentation operators applied to sequential recommendation. The original user interaction sequence is transformed through various operations including crop, mask, substitute, reorder, and noise addition to create augmented sequences that help improve model training and inference.}
    \label{fig:operators}
\end{figure}

To alleviate data sparsity, data augmentation is often used to augment the original data. As illustrated in Figure~\ref{fig:operators}, given an original sequence $s_u = [v_1, v_2, \ldots, v_{|s_u|}]$, we introduce eight representative augmentation operators that form the candidate set $\mathcal{A} = \{a_1, a_2, ..., a_K\}$ for our adaptive selection framework.

\begin{itemize}[leftmargin=*,noitemsep]
    \item \textbf{Crop:} Randomly select a continuous sub-sequence from $s_u$~\cite{zhou2020s3}:
    \begin{equation}
        s_u^a = \text{Aug}_{\text{crop}}(s_u) = [v_i, v_{i+1}, \ldots, v_{i+L-1}],
        \label{eq:crop}
    \end{equation}
    where $L$ is the length of the sub-sequence.
    
    \item \textbf{Reorder:} Randomly shuffle a continuous sub-sequence of $s_u$~\cite{zhou2020s3}:
    \begin{equation}
        s_u^a = \text{Aug}_{\text{reorder}}(s_u) = [v_1, \ldots, v'_i, \ldots, v'_{i+L-1}, \ldots, v_{|s_u|}],
        \label{eq:reorder}
    \end{equation}
    where $[v'_i, \ldots, v'_{i+L-1}]$ is a random permutation of $[v_i, \ldots, v_{i+L-1}]$.
    
    \item \textbf{Mask:} Randomly mask a proportion of items in $s_u$~\cite{sun2019bert4rec}:
    \begin{equation}
        s_u^a = \text{Aug}_{\text{mask}}(s_u) = [v'_1, v'_2, \ldots, v'_{|s_u|}],
        \label{eq:mask}
    \end{equation}
    where $v'_i$ will be replaced with the `[MASK]` token if $v_i$ is selected, otherwise $v'_i = v_i$.
    
    \item \textbf{Substitute:} Replace a proportion of items in $s_u$ with similar items~\cite{xie2022contrastive}:
    \begin{equation}
        s_u^a = \text{Aug}_{\text{substitute}}(s_u) = [v'_1, v'_2, \ldots, v'_{|s_u|}],
        \label{eq:substitute}
    \end{equation}
    where $v'_i$ is replaced with a correlated item based on representation similarity if $v_i$ is selected, otherwise $v'_i = v_i$.
    
    \item \textbf{Insert:} Insert a proportion of items into $s_u$~\cite{xie2022contrastive}:
    \begin{equation}
        s_u^a = \text{Aug}_{\text{insert}}(s_u) = [v_1, \ldots, v_i, v^c_i, \ldots, v_{|s_u|}],
        \label{eq:insert}
    \end{equation}
    where $v^c_i$ is an item correlated to the insertion position based on co-occurrence or representation similarity.
    
    \item \textbf{TNoise:} Inject uniform noise into the sequence representation~\cite{dang2025data}:
    \begin{equation}
        \mathbf{E}'_u = \text{Aug}_{\text{TNoise}}(\mathbf{E}_u) = \mathbf{E}_u + \boldsymbol{\epsilon}, \quad \boldsymbol{\epsilon} \sim \mathcal{U}(a, b),
        \label{eq:tnoise}
    \end{equation}
    where $\mathbf{E}_u \in \mathbb{R}^{|s_u| \times d}$ is the item representation matrix of the sequence $s_u$, $d$ is the representation dimension, and $a$, $b$ are hyperparameters controlling the noise magnitude.
    
    \item \textbf{TMask-B:} Set the representation of selected items to zero vectors, blocking them from model calculations~\cite{dang2025data}:
    \begin{equation}
        \mathbf{E}'_u = \text{Aug}_{\text{TMask-B}}(\mathbf{E}_u) = [e'_{v_1}, e'_{v_2}, \ldots, e'_{v_{|s_u|}}],
        \label{eq:tmask_b}
    \end{equation}
    where $e'_{v_i}$ will be set to a zero vector if $e_{v_i}$ is selected, otherwise $e'_{v_i} = e_{v_i}$.
    
    \item \textbf{TMask-R:} Directly remove the selected items without introducing any tokens~\cite{dang2025data}:
    \begin{equation}
        s_u^a = \text{Aug}_{\text{TMask-R}}(s_u) = [v_{i_1}, v_{i_2}, \ldots, v_{i_k}],
        \label{eq:tmask_r}
    \end{equation}
    where $\{i_1, i_2, \ldots, i_k\}$ are the indices of the items retained after removal.
\end{itemize}

These operators can be applied at test time to generate multiple augmented views of an input sequence. Recent studies show that operators such as TNoise and TMask can achieve competitive performance when used for test-time augmentation in sequential recommendation. However, the effectiveness of different operators may vary substantially across user sequences, questioning the validity of a uniform operator strategy.

\subsection{Test-Time Augmentation}
\label{sec:tta}

Test-time augmentation (TTA) augments the original inputs during model inference to generate multiple variants~\cite{kimura2021understanding}. Afterward, the final result is obtained by aggregating the model predictions on these variants~\cite{shanmugam2021better}. Unlike training-time augmentation which requires model retraining, TTA operates solely during inference without modifying the pretrained model, avoiding significant computational costs and operational overhead~\cite{zhang2022memo}.

Specifically, given an input sequence $s_u$ at test time, we can perform $m$ augmentations using an operator $a \in \mathcal{A}$ to obtain $m$ augmented sequences $\{\tilde{s}_u^{(i)}\}_{i=1}^m$, where $\tilde{s}_u^{(i)} = \text{Aug}(s_u, a)$. Finally, the model predicts the  next item $v^*$ based on average aggregation:
\begin{equation}
    v^* = \text{AverageAgg}\left(\sum_{i=1}^m f_{\Theta}(\tilde{s}_u^{(i)})\right),
    \label{eq:tta_aggregation}
\end{equation}
where $f_{\Theta}(\cdot)$ denotes the pretrained sequential recommendation model.

TTA has been shown to be effective across multiple domains, including computer vision~\cite{kim2020learning,shanmugam2020and} and graph learning~\cite{jin2022empowering}. 
Recently, TTA has also been introduced to sequential recommendation domain and shown to improve performance without retraining the underlying model~\cite{dang2025data}. Existing TTA methods, however, rely on a fixed augmentation operator that is applied uniformly to all test sequences. This naturally raises the following research question (Q1):
Does a uniform augmentation strategy remain optimal across heterogeneous user sequences in sequential recommendation? 


\section{EMPIRICAL STUDY}
\label{sec:empirical}
To answer this question, we conduct a systematic empirical study that examines the performance of different augmentation operators across user sequences with diverse characteristics. Specifically, we evaluate multiple test-time augmentation operators on user groups stratified by sequence properties, in order to assess whether operator effectiveness varies across different types of sequences. The experiments are conducted on the Amazon Beauty dataset~\cite{mcauley2015inferring} using a pretrained SASRec model~\cite{kang2018self} and eight commonly used augmentation operators \cite{dang2025data} introduced in the previous section.

\subsection{Length-based Analysis}
\label{sec:length_analysis}

We first investigate whether optimal augmentation operators vary across sequences of different lengths. We categorize users into three groups based on their sequence length: short (length $\leq$ 8), medium (8 $<$ length $\leq$ 20), and long (length $>$ 20). Table~\ref{tab:length_groups} presents the performance of all TTA operators on these three groups.

\noindent\textbf{Observation.} Different length groups prefer different optimal operators. Specifically, for short sequences, TMask-R performs best. For medium sequences, TNoise dominates. While for long sequences, Insert achieves the highest performance. No single operator is universally optimal across all groups.



\begin{table}[t]
    \centering
    \caption{Performance comparison of TTA operators across length-based user groups (Beauty dataset, SASRec).}
    \label{tab:length_groups}
    \small
    \resizebox{1.00\linewidth}{!}{
    \begin{tabular}{lcccccc|cc}
        \toprule
        \multirow{2}{*}{\textbf{Operator}} & \multicolumn{2}{c}{\textbf{Short}} & \multicolumn{2}{c}{\textbf{Medium}} & \multicolumn{2}{c}{\textbf{Long}} & \multicolumn{2}{c}{\textbf{Overall}} \\
        \cmidrule(lr){2-3} \cmidrule(lr){4-5} \cmidrule(lr){6-7} \cmidrule(lr){8-9}
        & H@10 & N@10 & H@10 & N@10 & H@10 & N@10 & H@10 & N@10 \\
        \midrule
        Base & 0.0553 & 0.0293 & 0.0685 & 0.0366 & 0.1197 & 0.0641 & 0.0584 & 0.0310 \\
        \midrule
        +Substitute & 0.0501 & 0.0262 & 0.0598 & 0.0306 & 0.0887 & 0.0439 & 0.0522 & 0.0271 \\
        +Crop & 0.0596 & 0.0313 & 0.0504 & 0.0253 & 0.0599 & 0.0329 & 0.0583 & 0.0305 \\
        +Reorder & 0.0627 & 0.0335 & 0.0649 & 0.0351 & 0.1042 & 0.0524 & 0.0638 & 0.0341 \\
        +Insert & 0.0569 & 0.0299 & 0.0698 & 0.0361 & \textbf{0.1242} & \textbf{0.0661} & 0.0601 & 0.0315 \\
        +TMask-R & \textbf{0.0632} & \textbf{0.0338} & 0.0649 & 0.0342 & 0.1153 & 0.0581 & \textbf{0.0645} & 0.0344 \\
        +TMask-B & 0.0394 & 0.0195 & \textbf{0.0730} & 0.0355 & 0.0887 & 0.0437 & 0.0451 & 0.0222 \\
        +TNoise & 0.0621 & 0.0335 & 0.0720 & \textbf{0.0375} & 0.1086 & 0.0575 & 0.0644 & \textbf{0.0346} \\
        \midrule
        \textbf{Oracle} & \multicolumn{2}{c}{\textit{+TMask-R}} & \multicolumn{2}{c}{\textit{+TNoise}} & \multicolumn{2}{c}{\textit{+Insert}} & \textbf{0.0659} & \textbf{0.0345} \\
        \bottomrule
    \end{tabular}
    }
\end{table}

\begin{table}[t]
    \centering
    \caption{Performance comparison of TTA operators across behavior-based user clusters with $K=3$ (Beauty dataset, SASRec).}
    \label{tab:cluster_k3}
    \small
    \resizebox{1.00\linewidth}{!}{
    \begin{tabular}{lcccccc|cc}
        \toprule
        \multirow{2}{*}{\textbf{Operator}} & \multicolumn{2}{c}{\textbf{Cluster 0}} & \multicolumn{2}{c}{\textbf{Cluster 1}} & \multicolumn{2}{c}{\textbf{Cluster 2}} & \multicolumn{2}{c}{\textbf{Overall}} \\
        \cmidrule(lr){2-3} \cmidrule(lr){4-5} \cmidrule(lr){6-7} \cmidrule(lr){8-9}
        & H@10 & N@10 & H@10 & N@10 & H@10 & N@10 & H@10 & N@10 \\
        \midrule
        Base & 0.0301 & 0.0160 & 0.1059 & 0.0577 & \textbf{0.1501} & \textbf{0.0778} & 0.0584 & 0.0310 \\
        \midrule
        +Substitute & 0.0253 & 0.0134 & 0.1169 & 0.0600 & 0.1145 & 0.0592 & 0.0522 & 0.0271 \\
        +Crop & 0.0338 & 0.0182 & \textbf{0.1285} & 0.0675 & 0.0944 & 0.0503 & 0.0575 & 0.0306 \\
        +Reorder & 0.0352 & 0.0186 & 0.1239 & 0.0664 & 0.1344 & 0.0716 & 0.0630 & 0.0335 \\
        +Insert & 0.0318 & 0.0168 & 0.1191 & 0.0646 & 0.1415 & 0.0732 & 0.0607 & 0.0321 \\
        +TMask-R & \textbf{0.0366} & \textbf{0.0197} & 0.1282 & 0.0682 & 0.1337 & 0.0703 & \textbf{0.0645} & 0.0344 \\
        +TMask-B & 0.0183 & 0.0089 & 0.0863 & 0.0406 & 0.1368 & 0.0705 & 0.0451 & 0.0222 \\
        +TNoise & 0.0350 & 0.0188 & 0.1247 & \textbf{0.0684} & 0.1436 & 0.0758 & 0.0641 & \textbf{0.0345} \\
        \midrule
        \textbf{Oracle} & \multicolumn{2}{c}{\textit{+TMask-R}} & \multicolumn{2}{c}{\textit{+Crop}} & \multicolumn{2}{c}{\textit{Base}} & \textbf{0.0673} & \textbf{0.0354} \\
        \bottomrule
    \end{tabular}}
\end{table}

\begin{table*}[t]
    \centering
    \caption{Performance comparison of TTA operators across behavior-based user clusters with $K=5$ (Beauty dataset, SASRec).}
    \label{tab:cluster_k5}
    \small
    \begin{tabular}{lcccccccccc|cc}
        \toprule
        \multirow{2}{*}{\textbf{Operator}}  & \multicolumn{2}{c}{\textbf{Cluster 0}} & \multicolumn{2}{c}{\textbf{Cluster 1}} & \multicolumn{2}{c}{\textbf{Cluster 2}} & \multicolumn{2}{c}{\textbf{Cluster 3}} & \multicolumn{2}{c}{\textbf{Cluster 4}} & \multicolumn{2}{c}{\textbf{Overall}} \\
        \cmidrule(lr){2-3} \cmidrule(lr){4-5} \cmidrule(lr){6-7} \cmidrule(lr){8-9} \cmidrule(lr){10-11} \cmidrule(lr){12-13}
        & H@10 & N@10 & H@10 & N@10 & H@10 & N@10 & H@10 & N@10 & H@10 & N@10 & H@10 & N@10 \\
        \midrule
        Base & 0.0302 & 0.0166 & 0.1019 & 0.0516 & 0.1102 & 0.0562 & \textbf{0.3580} & \textbf{0.2128} & 0.0364 & 0.0183 & 0.0584 & 0.0310 \\
        \midrule
        +Substitute & 0.0203 & 0.0107 & 0.1208 & 0.0594 & 0.0948 & 0.0483 & 0.2218 & 0.1268 & 0.0402 & 0.0217 & 0.0522 & 0.0271 \\
        +Crop & 0.0297 & 0.0165 & \textbf{0.1257} & 0.0618 & 0.0436 & 0.0228 & 0.3277 & 0.1975 & 0.0499 & 0.0258 & 0.0575 & 0.0306 \\
        +Reorder & \textbf{0.0336} & 0.0179 & 0.1178 & 0.0587 & 0.0899 & 0.0488 & 0.3513 & 0.2098 & 0.0478 & 0.0246 & 0.0630 & 0.0335 \\
        +Insert & 0.0294 & 0.0158 & 0.1171 & 0.0603 & \textbf{0.1124} & 0.0570 & 0.3008 & 0.1770 & 0.0435 & 0.0227 & 0.0607 & 0.0321 \\
        +TMask-R & 0.0335 & \textbf{0.0185} & 0.1224 & 0.0618 & 0.0886 & 0.0462 & 0.3529 & 0.2119 & \textbf{0.0517} & \textbf{0.0260} & \textbf{0.0645} & 0.0344 \\
        +TMask-B & 0.0163 & 0.0083 & 0.0874 & 0.0406 & 0.1106 & \textbf{0.0573} & 0.2622 & 0.1356 & 0.0276 & 0.0128 & 0.0451 & 0.0222 \\
        +TNoise & 0.0328 & 0.0178 & 0.1211 & \textbf{0.0626} & 0.1080 & 0.0569 & 0.3328 & 0.2009 & 0.0468 & 0.0244 & 0.0641 & \textbf{0.0345} \\
        \midrule
        \textbf{Oracle} & \multicolumn{2}{c}{\textit{+Reorder}} & \multicolumn{2}{c}{\textit{+Crop}} & \multicolumn{2}{c}{\textit{+Insert}} & \multicolumn{2}{c}{\textit{Base}} & \multicolumn{2}{c}{\textit{+TMask-R}} & \textbf{0.0679} & \textbf{0.0361} \\
        \bottomrule
    \end{tabular}
\end{table*}

\subsection{Behavior-based Clustering Analysis}
\label{sec:behavior_analysis}

To examine whether operator preferences are driven by more than simple sequence length effects, we further cluster users based on their behavioral patterns, represented by sequence embeddings, using K-means clustering method~\cite{arthur2006k}. We then repeat the above analysis within each cluster. Tables~\ref{tab:cluster_k3} and \ref{tab:cluster_k5} present results for $K=3$ and $K=5$ clusters respectively.

\noindent\textbf{Observation.} 
Behavior-based clusters also exhibit distinct operator preferences. For instance, with $K=3$, Cluster 0 benefits most from TMask-R, while Cluster 1 performs best with Crop, indicating clear heterogeneity in operator effectiveness across behavioral groups. Similar patterns persist under finer-grained clustering ($K=5$). Taken together, Tables~\ref{tab:cluster_k3} and \ref{tab:cluster_k5} provide consistent empirical evidence that optimal augmentation operators are behavior-dependent.


\subsection{Potential of Adaptive Selection}
\label{sec:personalized_potential}

Having established that optimal operators vary by group and a fixed and uniform augmentation strategy cannot adapt to the intrinsic characteristics of all user sequences, we quantify the potential gain from adaptive selection. To this end, we compare two strategies: a fixed strategy applying the same best single operator to all users, and a personalized Oracle strategy assigning the optimal operator to each group.

\noindent\textbf{Observation.} The personalized Oracle strategy achieves consistent improvements over the best fixed strategy. The improvement is more pronounced with behavior-based clustering and increases progressively with finer cluster granularity (from $K=3$ to $K=5$).

\noindent\textbf{Analysis.} The results indicate that more informative user representations enable more precise personalized augmentation. Compared with simple length-based grouping, behavior-based clustering better captures latent factors that influence augmentation effectiveness. Moreover, the consistent performance improvements observed with an increasing number of clusters suggest that finer-grained personalization may further enhance TTA performance. This trend implies that, in principle, assigning augmentation operators at the individual sequence level may unlock additional performance gains.

\begin{figure*}[t]
    \centering
    \includegraphics[width=0.99\textwidth]{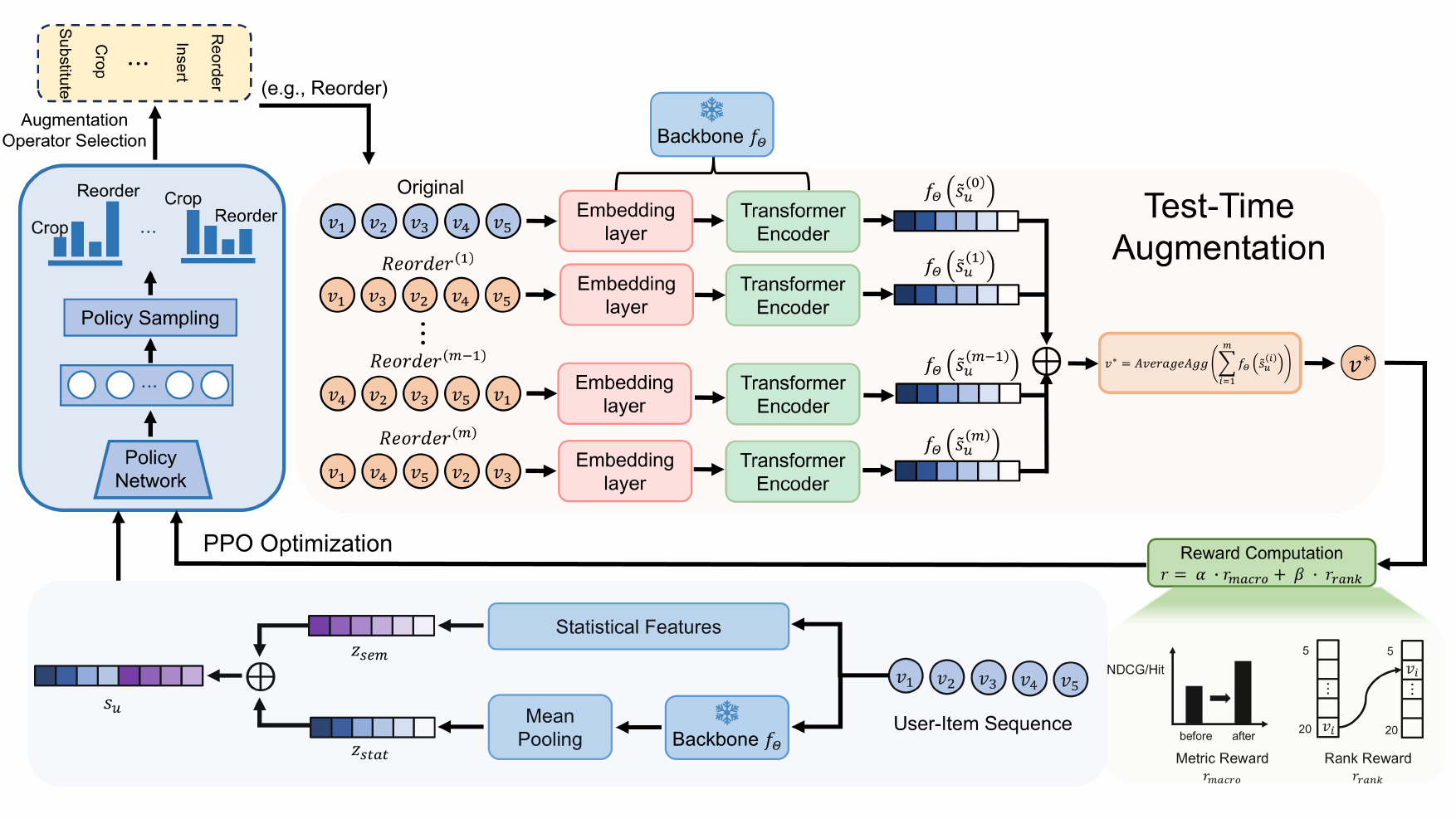}
    \caption{Overall framework of AdaTTA. (1) The input user sequence is encoded into semantic and statistical features to form the state representation. (2) The Actor-Critic policy network selects the optimal augmentation operator. (3) The selected operator generates augmented sequences for inference. (4) The reward function evaluates the recommendation performance to guide policy optimization.}
    \Description{The AdaTTA framework consists of four main components: state representation with semantic and statistical views, an Actor-Critic policy network for operator selection, test-time augmentation and inference, and a reward-based optimization loop.}
    \label{fig:framework}
\end{figure*}

\subsection{Summary and Implications}

Based on the above analysis, we summarize our findings as follows:

\begin{itemize}[leftmargin=*,noitemsep]
    \item \textbf{No universal best operator:} The effectiveness of augmentation operators is highly dependent on sequence characteristics like length and behavior patterns.
    \item \textbf{Personalized strategy holds significant potential:} Even based on simple groupings, personalized selection yields noticeable improvements over fixed strategies.
\end{itemize}

These findings provide an affirmative answer to our first research question (Q1) and reveal a more fundamental challenge: while adaptive selection at the group level yields clear benefits, our empirical analysis still relies on uniform augmentation strategies within each group. Ideally, augmentation operators should be selected dynamically for each individual user sequence. This observation motivates our second research question (Q2) and the AdaTTA framework introduced in the next section, which aims to enable fully adaptive, sequence-level operator selection.

\section{METHODOLOGY}

In this section, we present AdaTTA, an adaptive test-time augmentation framework for sequential recommendation. Figure~\ref{fig:framework} illustrates the overall architecture. Specifically, We formulate test-time augmentation as a sequential decision-making problem, in which an agent dynamically selects augmentation operators for each input user sequence with the objective of maximizing recommendation performance. To solve this problem, we employ reinforcement learning to learn an operator selection policy, optimized using a macro-rank reward design.

\subsection{Problem Formulation}
Let $\mathcal{U}$ and $\mathcal{I}$ denote the sets of users and items. A user's historical interaction sequence is represented as $S_u = [v_1, v_2, \dots, v_t]$. Given a pre-trained backbone recommender $f_{\Theta}$, our goal is to learn a policy $\pi_\phi(a|S_u)$ that selects an optimal augmentation action $a \in \mathcal{A}$ to generate an enhanced view $S'_u$, such that the prediction accuracy of $f_{\Theta}$ on $S'_u$ is maximized.

\subsection{Multi-View State Representation}

To facilitate robust decision-making, we construct a state vector $\mathbf{s}_u$ that captures both the semantic preferences and structural characteristics of the user sequence. Specifically, the state representation fuses two complementary views:
\begin{equation}
    \mathbf{s}_u = \mathbf{z}_{sem} \oplus \mathbf{z}_{stat}
\end{equation}
\noindent where $\mathbf{z}_{sem} \in \mathbb{R}^d$ represents the semantic view, obtained by mean-pooling the hidden representations from the backbone recommendation model, and $\mathbf{z}_{stat} \in \mathbb{R}^k$ denotes the statistical view, which explicitly encodes structural properties such as sequence length, diversity, and noise levels. This dual-view hybrid state design reflects our empirical observation that both structural properties (\textit{e.g.}, sequence length) and semantic behavior patterns play critical roles in determining operator effectiveness.


\subsection{Actor-Critic Policy Learning}

We employ an Actor-Critic architecture~\cite{konda1999actor} to parameterize the operator selection policy. The network consists of a shared encoder to extract high-level representations from the hybrid state vector $\mathbf{s}_u$, followed by two specific heads:

\textbf{Actor Head} outputs a categorical distribution $\pi_\phi(a|\mathbf{s}_u)$ over the discrete action space $\mathcal{A}$ (\textit{e.g.}, \textit{Crop}, \textit{Substitute}, \textit{Insert}).

\textbf{Critic Head} estimates the value function $V_\psi(\mathbf{s}_u)$ to guide the optimization process.

During inference, the selected action $a$ triggers a specific transformation $T_a$, generating $m$ augmented sequence samples to stabilize the next item prediction. Notably, the action space is intentionally lightweight, enabling efficient policy learning.

\subsection{Reward Design and Optimization}
Designing a dense and effective reward signal is critical for reinforcement learning in ranking tasks~\cite{chen2019top}. We propose a multi-objective reward function that balances global metric improvement with fine-grained ranking stability.

\subsubsection{Objective Function}
The reward $r_t$ for a given action is defined as a weighted combination of two components:
\begin{equation}
    r_t = \alpha \cdot r_{macro} + \beta \cdot r_{rank}
\end{equation}
\noindent where $\alpha$ and $\beta$ are hyperparameters balancing the two objectives.

\textbf{Macro Reward ($r_{macro}$)} reflects the improvement in top-$K$ evaluation metrics. To address reward sparsity, we implement a \textit{tiered threshold mechanism}, assigning graded positive rewards for significant metric gains and penalties for performance degradation.

\textbf{Rank Reward ($r_{rank}$)} captures the absolute position shift of the target item in the recommendation list. It provides granular supervision even when the target item remains outside the top-$K$ cutoff, encouraging the model to progressively improve the ranking of correct items.

\subsubsection{Optimization Strategy}
We optimize the policy using Proximal Policy Optimization~\cite{schulman2017proximal}. PPO improves training stability by clipping the probability ratio of policy updates. The final loss function incorporates an entropy regularization term to prevent premature convergence:
\begin{equation}
    \mathcal{L} = \mathcal{L}_{CLIP}(\phi) - c_1 \mathcal{L}_{VF}(\psi) + c_2 \mathcal{H}(\pi_\phi)
\end{equation}
\noindent where $\mathcal{L}_{CLIP}$ is the surrogate objective of PPO, $\mathcal{L}_{VF}$ is the value function loss, and $\mathcal{H}$ denotes the entropy of the policy. The coefficients $c_2$ and the augmentation budget $m$ are analyzed in the parameter analysis section.

\section{EXPERIMENTS}
\label{sec:experiments}

\begin{table*}[t]
    \centering
    \caption{Performance comparison of different test-time augmentation methods. 
    The best performance is \textbf{bolded} and the second best is \underline{underlined}. 
    The 'Improve-$\alpha$' and 'Improve-$\beta$' represent the relative improvements of our best result over the based model and the best baseline method, respectively.}
    \label{tab:main_results}
    {\fontsize{8.3pt}{8.8pt}\selectfont
    \setlength{\tabcolsep}{4pt}
    \begin{tabular}{l|cccccc|cccccc}
        \toprule
        \multicolumn{1}{c|}{\multirow{2}{*}{Method}} & \multicolumn{6}{c|}{Beauty} & \multicolumn{6}{c}{Sports} \\
        & H@5 & N@5 & H@10 & N@10 & H@20 & N@20 & H@5 & N@5 & H@10 & N@10 & H@20 & N@20 \\
        \midrule
        GRU4Rec (Base) & 0.0204 & 0.0125 & 0.0376 & 0.0180 & 0.0652 & 0.0249 & 0.0083 & 0.0051 & 0.0157 & 0.0075 & 0.0291 & 0.0108 \\
        \midrule
        + Substitute & 0.0244 & 0.0148 & 0.0441 & 0.0211 & 0.0732 & 0.0284 & 0.0138 & 0.0087 & 0.0245 & 0.0121 & 0.0424 & 0.0166 \\
        + Crop & 0.0271 & 0.0173 & 0.0433 & 0.0225 & 0.0700 & 0.0292 & 0.0154 & 0.0096 & \underline{0.0274} & 0.0135 & 0.0453 & 0.0180 \\
        + Reorder & 0.0263 & 0.0161 & 0.0443 & 0.0219 & 0.0724 & 0.0289 & 0.0151 & 0.0092 & 0.0262 & 0.0127 & 0.0453 & 0.0175 \\
        + Insert & 0.0245 & 0.0148 & 0.0432 & 0.0208 & 0.0739 & 0.0285 & 0.0140 & 0.0087 & 0.0255 & 0.0123 & 0.0447 & 0.0172 \\
        + TNoise & 0.0204 & 0.0123 & 0.0378 & 0.0178 & 0.0637 & 0.0243 & 0.0079 & 0.0047 & 0.0156 & 0.0071 & 0.0279 & 0.0102 \\
        + TMask-B & 0.0251 & 0.0155 & 0.0433 & 0.0213 & 0.0675 & 0.0273 & \underline{0.0156} & 0.0099 & 0.0263 & 0.0133 & 0.0441 & 0.0178 \\
        + TMask-R & \underline{0.0287} & \underline{0.0178} & \underline{0.0474} & \underline{0.0238} & \underline{0.0744} & \underline{0.0306} & \underline{0.0156} & \underline{0.0100} & 0.0273 & \underline{0.0137} & \underline{0.0461} & \underline{0.0184} \\
        \midrule
        AdaTTA & \textbf{0.0295} & \textbf{0.0182} & \textbf{0.0492} & \textbf{0.0245} & \textbf{0.0797} & \textbf{0.0322} & \textbf{0.0186} & \textbf{0.0118} & \textbf{0.0311} & \textbf{0.0158} & \textbf{0.0511} & \textbf{0.0208} \\
        Improve-$\alpha$ & 44.60\% & 45.60\% & 30.85\% & 36.11\% & 22.23\% & 29.31\% & 124.09\% & 131.37\% & 98.08\% & 110.66\% & 75.60\% & 92.59\% \\
        Improve-$\beta$ & 2.78\% & 2.24\% & 3.79\% & 2.94\% & 7.12\% & 5.22\% & 19.23\% & 18.00\% & 13.50\% & 15.32\% & 10.84\% & 13.04\% \\
        \midrule
        \midrule
        SASRec (Base) & 0.0373 & 0.0243 & 0.0584 & 0.0310 & 0.0914 & 0.0393 & 0.0190 & 0.0121 & 0.0317 & 0.0161 & 0.0495 & 0.0206 \\
        \midrule
        + Substitute & 0.0376 & 0.0239 & 0.0597 & 0.0309 & 0.0950 & 0.0399 & 0.0212 & 0.0126 & 0.0368 & 0.0177 & 0.0590 & 0.0232 \\
        + Crop & 0.0399 & 0.0256 & 0.0639 & 0.0333 & 0.0954 & 0.0412 & \underline{0.0262} & 0.0167 & 0.0413 & 0.0215 & 0.0638 & 0.0272 \\
        + Reorder & 0.0411 & 0.0267 & 0.0648 & 0.0343 & \underline{0.1003} & 0.0432 & 0.0261 & 0.0166 & 0.0416 & 0.0215 & 0.0643 & 0.0272 \\
        + Insert & 0.0375 & 0.0241 & 0.0599 & 0.0313 & 0.0937 & 0.0398 & 0.0207 & 0.0126 & 0.0358 & 0.0175 & 0.0566 & 0.0227 \\
        + TNoise & \underline{0.0418} & 0.0272 & 0.0656 & 0.0348 & 0.0985 & 0.0431 & 0.0239 & 0.0152 & 0.0394 & 0.0202 & 0.0600 & 0.0253 \\
        + TMask-B & 0.0377 & 0.0244 & 0.0590 & 0.0312 & 0.0929 & 0.0398 & 0.0205 & 0.0129 & 0.0343 & 0.0173 & 0.0542 & 0.0223 \\
        + TMask-R & 0.0416 & \underline{0.0273} & \underline{0.0657} & \underline{0.0350} & 0.0996 & \underline{0.0435} & \underline{0.0262} & \underline{0.0168} & \underline{0.0428} & \underline{0.0222} & \underline{0.0660} & \underline{0.0280} \\
        \midrule
        AdaTTA & \textbf{0.0438} & \textbf{0.0281} & \textbf{0.0692} & \textbf{0.0363} & \textbf{0.1050} & \textbf{0.0453} & \textbf{0.0282} & \textbf{0.0181} & \textbf{0.0444} & \textbf{0.0233} & \textbf{0.0677} & \textbf{0.0292} \\
        Improve-$\alpha$ & 17.42\% & 15.63\% & 18.49\% & 17.09\% & 14.87\% & 15.26\% & 48.42\% & 49.58\% & 40.06\% & 44.72\% & 36.76\% & 41.74\% \\
        Improve-$\beta$ & 4.78\% & 2.93\% & 5.32\% & 3.71\% & 4.68\% & 4.13\% & 7.63\% & 7.73\% & 3.73\% & 4.95\% & 2.57\% & 4.28\% \\
        \bottomrule
    \end{tabular}

    \vspace{0.3cm}
    
    \begin{tabular}{l|cccccc|cccccc}
        \toprule
        \multicolumn{1}{c|}{\multirow{2}{*}{Method}} & \multicolumn{6}{c|}{Home} & \multicolumn{6}{c}{Yelp} \\
        & H@5 & N@5 & H@10 & N@10 & H@20 & N@20 & H@5 & N@5 & H@10 & N@10 & H@20 & N@20 \\
        \midrule
        GRU4Rec (Base) & 0.0035 & 0.0022 & 0.0063 & 0.0031 & 0.0116 & 0.0044 & 0.0082 & 0.0048 & 0.0158 & 0.0072 & 0.0298 & 0.0108 \\
        \midrule
        + Substitute & 0.0051 & 0.0031 & 0.0095 & 0.0045 & 0.0175 & 0.0065 & 0.0111 & 0.0066 & 0.0203 & 0.0096 & 0.0364 & 0.0136 \\
        + Crop & 0.0058 & 0.0035 & 0.0108 & 0.0051 & 0.0199 & 0.0074 & 0.0111 & 0.0067 & 0.0202 & 0.0096 & 0.0351 & 0.0134 \\
        + Reorder & 0.0053 & 0.0033 & 0.0103 & 0.0049 & 0.0195 & 0.0072 & 0.0101 & 0.0061 & 0.0190 & 0.0089 & 0.0349 & 0.0129 \\
        + Insert & 0.0056 & 0.0034 & 0.0104 & 0.0050 & 0.0187 & 0.0071 & 0.0111 & 0.0066 & 0.0206 & 0.0096 & \underline{0.0366} & 0.0136 \\
        + TNoise & 0.0036 & 0.0023 & 0.0061 & 0.0030 & 0.0110 & 0.0043 & 0.0072 & 0.0043 & 0.0144 & 0.0066 & 0.0275 & 0.0098 \\
        + TMask-B & 0.0053 & 0.0031 & 0.0092 & 0.0044 & 0.0176 & 0.0065 & 0.0112 & \underline{0.0069} & 0.0202 & 0.0097 & 0.0362 & \underline{0.0138} \\
        + TMask-R & \underline{0.0063} & \underline{0.0038} & \underline{0.0118} & \underline{0.0056} & \underline{0.0212} & \underline{0.0080} & \underline{0.0114} & \underline{0.0069} & \underline{0.0207} & \underline{0.0099} & 0.0364 & \underline{0.0138} \\
        \midrule
        AdaTTA & \textbf{0.0079} & \textbf{0.0048} & \textbf{0.0133} & \textbf{0.0066} & \textbf{0.0229} & \textbf{0.0090} & \textbf{0.0119} & \textbf{0.0071} & \textbf{0.0217} & \textbf{0.0103} & \textbf{0.0381} & \textbf{0.0144} \\
        Improve-$\alpha$ & 125.71\% & 118.18\% & 111.11\% & 112.90\% & 97.41\% & 104.54\% & 45.12\% & 47.91\% & 37.34\% & 43.05\% & 27.85\% & 33.33\% \\
        Improve-$\beta$ & 25.39\% & 26.31\% & 12.71\% & 17.85\% & 8.01\% & 12.5\% & 4.38\% & 2.89\% & 4.83\% & 4.04\% & 4.09\% & 4.34\% \\
        \midrule
        \midrule
        SASRec (Base) & 0.0097 & 0.0061 & 0.0160 & 0.0081 & 0.0249 & 0.0104 & 0.0142 & 0.0088 & 0.0253 & 0.0124 & 0.0430 & 0.0168 \\
        \midrule
        + Substitute & 0.0104 & 0.0066 & 0.0173 & 0.0088 & 0.0283 & 0.0116 & 0.0145 & 0.0090 & 0.0258 & 0.0126 & 0.0446 & 0.0173 \\
        + Crop & 0.0110 & 0.0069 & 0.0180 & 0.0092 & 0.0289 & 0.0119 & 0.0162 & 0.0100 & 0.0284 & 0.0140 & 0.0485 & 0.0190 \\
        + Reorder & 0.0116 & 0.0073 & \underline{0.0195} & 0.0098 & 0.0304 & 0.0126 & 0.0164 & 0.0101 & 0.0286 & 0.0140 & 0.0482 & 0.0189 \\
        + Insert & 0.0109 & 0.0068 & 0.0181 & 0.0091 & 0.0285 & 0.0117 & 0.0145 & 0.0090 & 0.0257 & 0.0126 & 0.0437 & 0.0171 \\
        + TNoise & 0.0113 & 0.0071 & 0.0191 & 0.0096 & 0.0311 & 0.0126 & 0.0162 & 0.0102 & 0.0278 & 0.0140 & 0.0462 & 0.0186 \\
        + TMask-B & 0.0099 & 0.0064 & 0.0164 & 0.0084 & 0.0261 & 0.0108 & 0.0154 & 0.0096 & 0.0268 & 0.0132 & 0.0460 & 0.0180 \\
        + TMask-R & \underline{0.0117} & \underline{0.0075} & 0.0194 & \underline{0.0099} & \underline{0.0313} & \underline{0.0129} & \underline{0.0171} & \underline{0.0106} & \underline{0.0297} & \underline{0.0146} & \underline{0.0499} & \underline{0.0196} \\
        \midrule
        AdaTTA & \textbf{0.0124} & \textbf{0.0077} & \textbf{0.0205} & \textbf{0.0103} & \textbf{0.0324} & \textbf{0.0133} & \textbf{0.0174} & \textbf{0.0107} & \textbf{0.0299} & \textbf{0.0147} & \textbf{0.0503} & \textbf{0.0198} \\
        Improve-$\alpha$ & 27.83\% & 26.22\% & 28.12\% & 27.16\% & 30.12\% & 27.88\% & 22.53\% & 21.59\% & 18.18\% & 18.54\% & 16.97\% & 17.85\% \\
        Improve-$\beta$ & 5.98\% & 2.66\% & 5.12\% & 4.04\% & 3.51\% & 3.10\% & 1.75\% & 0.94\% & 0.67\% & 0.68\% & 0.80\% & 1.02\% \\
        \bottomrule
    \end{tabular}}
\end{table*}

\subsection{Experiment Setup}

\subsubsection{Datasets}
The experiments are conducted on four datasets. Beauty, Sports, and Home are obtained from Amazon~\cite{mcauley2015inferring} and correspond to the ''Beauty'', ''Sports and Outdoors'', and ''Home'' categories, respectively. Yelp is a business dataset, which we use its first version released in 2018 ~\cite{dang2023uniform,liu2021contrastive}. Data preprocessing follows the standard protocol adopted in prior work~\cite{dang2025data}.


\subsubsection{Baselines}
In our experiments, we adopt pre-trained SASRec~\cite{kang2018self} and GRU4Rec~\cite{hidasi2015session}  models augmented with a single test-time augmentation (TTA) operator as comparative baselines. Following prior work~\cite{dang2025data}, each baseline applies a fixed augmentation operator uniformly across all test sequences. This experimental setup allows us to systematically evaluate the benefits of adaptive operator selection by comparing against representative fixed-TTA strategies across different datasets.

\subsubsection{Evaluation Settings}
\label{subsubsec:eval-settings}
We adopt the leave-one-out strategy to partition sequences into training, validation, and test sets. We rank the prediction over the whole item set rather than negative sampling, otherwise leading to biased discoveries~\cite{krichene2020sampled}. The evaluation metrics include Hit Ratio@K (H@K) and Normalized Discounted Cumulative Gain@K (N@K). We report results with $K \in \{5,10,20\}$. Generally, greater values imply better ranking performance.

\subsubsection{Implementation Details}

For all baselines, we adopt the implementations provided by the authors and carefully tune hyperparameters as reported in the original papers. We set the embedding size to 64 and the batch size to 256, using Adam optimizer with learning rate of 0.001. For AdaTTA, we train the policy network using PPO with policy batch size of 512, discount factor $\gamma=0.96$, value loss coefficient 0.25, clip parameter $\epsilon=0.2$, and gradient clipping norm 1.0. We set the number of TTA augmentations $m=10$. All experiments are repeated five times, and we report the average results across five runs for all methods.


\subsection{Performance Comparison}
\label{sec:comparison_tta}

We compare the performance of the base models with different test-time augmentation methods, and the results are summarized in Table~\ref{tab:main_results}. We have the following observations and conclusions:

(1) The effectiveness of different fixed augmentation strategies varies significantly, and their superiority changes across models and datasets, which confirms the core finding from our empirical study. For GRU4Rec, TMask-R performs best in most cases, \textit{e.g.}, improving H@10 from 0.0376 to 0.0474 on the Beauty dataset. However, for SASRec, there is no single absolute winner, with TNoise, Reorder, and TMask-R each excelling on different metrics. This indicates the difficulty of finding a universally “best” fixed strategy.

(2) Our proposed AdaTTA framework consistently outperforms all fixed-strategy baselines across all experimental settings, providing a strong affirmative answer to our second research question (Q2). AdaTTA not only significantly surpasses the base models but also stably exceeds the best fixed strategy in each setting. For instance, the most impressive gain is observed for GRU4Rec on the Sports dataset: AdaTTA improves H@5 by 124.09\% over the base model and achieves an additional 19.23\% gain over the best fixed strategy (TMask-R) in that scenario. This result unequivocally proves that developing a learning-based framework to automatically select optimal operators based on sequence characteristics is both feasible and highly effective.


\subsection{Ablation Study}

\begin{table}[t]
    \centering
    \caption{Ablation study of the action space.}
    \label{tab:action_ablation}
    \begin{tabular}{lcccc}
        \toprule
        \multicolumn{1}{c}{\multirow{2}{*}{\textbf{Configuration}}} & \multicolumn{2}{c}{\textbf{Beauty}} & \multicolumn{2}{c}{\textbf{Home}} \\
        \cmidrule(lr){2-3} \cmidrule(lr){4-5}
         & H@5 & H@10 & H@5 & H@10 \\
        \midrule
        5-Action & 0.0421 & 0.0656 & 0.0118 & 0.0189 \\
        6-Action (+TMask-R) & 0.0425 & 0.0663 & 0.0119 & 0.0193 \\
        7-Action (+TMask-B) & 0.0428 & 0.0688 & 0.0114 & 0.0196 \\
        8-Action (+TNoise) & \textbf{0.0438} & \textbf{0.0692} & \textbf{0.0124} & \textbf{0.0205} \\
        \bottomrule
    \end{tabular}
\end{table}

\noindent\textbf{Action Space Ablation.} We investigate the contribution of different augmentation strategies by progressively expanding the action space. Table~\ref{tab:action_ablation} presents the results on the Beauty and Home datasets with the SASRec backbone. On Beauty, performance improves consistently as the action space expands. Performance plateaus when the action space reaches 7 and 8 strategies, suggesting that including all strategies does not yield additional benefit. On Home, the results show more variability, with a moderate action space achieving the best performance. Overall, these findings suggest that while a richer action space is generally beneficial, adaptive operator selection mitigates the need for excessive strategy diversity; even a compact action space (\textit{e.g.}, 5 simple actions) is sufficient to achieve strong and robust performance.

\begin{table}[t]
    \centering
    \caption{Ablation study of the reward design.}
    \label{tab:reward_ablation}
    \begin{tabular}{lcccccc}
        \toprule
        \multicolumn{1}{c}{\multirow{2}{*}{\textbf{Configuration}}} & \multicolumn{2}{c}{\textbf{Beauty}} & \multicolumn{2}{c}{\textbf{Home}} \\
        \cmidrule(lr){2-3} \cmidrule(lr){4-5}
        & H@5 & H@10 & H@5 & H@10 \\
        \midrule
        Metric Only & 0.0414 & 0.0647 & 0.0109 & 0.0187 \\
        Rank Only & 0.0436 & 0.0664 & 0.0117 & 0.0197 \\
        \midrule
        \textbf{AdaTTA} & \textbf{0.0438} & \textbf{0.0692} & \textbf{0.0124} & \textbf{0.0205} \\
        \bottomrule
    \end{tabular}
\end{table}

\noindent\textbf{Reward Design Ablation.} We evaluate the contribution of different reward components by comparing configurations using metric-only reward, rank-only reward, and their combinations. The ablation results on Beauty and Home datasets with the SASRec backbone are presented in Table~\ref{tab:reward_ablation}. The configuration combining both reward components consistently outperforms those using either component alone. 

\begin{table}[t]
    \centering
    \caption{Ablation study of hybrid state representation.}
    \label{tab:feature_ablation}
    \begin{tabular}{lcccccc}
        \toprule
        \multicolumn{1}{c}{\multirow{2}{*}{\textbf{Configuration}}} & \multicolumn{2}{c}{\textbf{Beauty}} & \multicolumn{2}{c}{\textbf{Home}} \\
        \cmidrule(lr){2-3} \cmidrule(lr){4-5}
        & H@5 & H@10 & H@5 & H@10 \\
        \midrule
        Embedding Only & 0.0427 & 0.0674 & 0.0122 & 0.0200 \\
        Stats Only & 0.0423 & 0.0661 & 0.0120 & 0.0195 \\
        \midrule
        \textbf{AdaTTA} & \textbf{0.0438} & \textbf{0.0692} & \textbf{0.0124} & \textbf{0.0205} \\
        \bottomrule
    \end{tabular}
\end{table}

\noindent\textbf{State Representation Ablation.} We evaluate the contribution of different components in the state representation by comparing configurations using only embedding features, only statistical features, and their combination. The ablation results on the Beauty and Home datasets with the SASRec backbone are presented in Table~\ref{tab:feature_ablation}. The configuration combining both feature types consistently outperforms those using either component alone. This result suggests that semantic embeddings and statistical features provide complementary signals to the policy network: embeddings encode user preference semantics, while statistical features capture structural properties of interaction sequences (\textit{e.g.}, length and diversity) that are both informative for effective operator selection.

\subsection{Efficiency Analysis}

\begin{table}[t]
\centering
\caption{Inference time comparison of different test-time augmentation methods.}
\label{tab:efficiency_analysis}
\resizebox{\columnwidth}{!}{%
\begin{tabular}{lcccc}
\toprule


\multicolumn{1}{c}{\multirow{2}{*}{\textbf{Method}}} & \multicolumn{2}{c}{\textbf{Beauty}} & \multicolumn{2}{c}{\textbf{Home}} \\
\cmidrule(lr){2-3} \cmidrule(lr){4-5}
 & \textbf{Inf.Time(min)} & \textbf{Ratio} & \textbf{Inf.Time(min)} & \textbf{Ratio} \\
\midrule
+ Substitute  & 0.28 & 1.07× & 0.89 & 1.05× \\
+ Crop        & 0.18 & 1.67× & 0.55 & 1.69× \\
+ Reorder     & 0.19 & 1.58× & 0.63 & 1.48× \\
+ Insert      & 0.24 & 1.25× & 1.12 & 0.83× \\
+ TNoise      & 0.09 & 3.33× & 0.34 & 2.74× \\
+ TMask-B     & 0.18 & 1.67× & 0.59 & 1.58× \\
+ TMask-R     & 0.18 & 1.67× & 0.63 & 1.48× \\
\midrule
\textbf{AdaTTA} & 0.30 & -- & 0.93 & -- \\
\bottomrule
\end{tabular}
}
\end{table}


\noindent\textbf{Efficiency Analysis.} Table~\ref{tab:efficiency_analysis} compares the inference time of different test-time augmentation methods over all test sequences with the SASRec backbone. 
Compared to fixed augmentation strategies, AdaTTA incurs a slightly higher inference time as it additionally performs policy-based operator selection. However, this overhead remains limited—for example, on the Home dataset, AdaTTA takes 0.93 minutes, which is only 1.05× slower than Substitute and 1.48× slower than TMask-R (the best baseline). Given the substantial performance gains achieved by AdaTTA (Table~\ref{tab:main_results}), this moderate increase in inference time represents a favorable trade-off. In practical recommender system deployments, even small improvements in recommendation quality can translate into significant commercial value, making the additional computational cost acceptable. Moreover, the policy network can be further optimized at deployment time using standard techniques such as model quantization to reduce latency.

\subsection{Parameter Analysis}

\begin{table}[t]
\centering
\caption{Performance with different entropy coefficients $c_2$.}
\label{tab:entropy_ablation}
\resizebox{\columnwidth}{!}{%
\begin{tabular}{lccccccc}
\toprule
\multirow{2}{*}{\textbf{Dataset}} & \multirow{2}{*}{$c_2$} & \multicolumn{3}{c}{\textbf{HIT}} & \multicolumn{3}{c}{\textbf{NDCG}} \\
\cmidrule(lr){3-5} \cmidrule(lr){6-8}
& & @5 & @10 & @20 & @5 & @10 & @20 \\
\midrule
\multirow{5}{*}{Beauty} 
& 0.01 & 0.0419 & 0.0663 & 0.1003 & 0.0272 & 0.0351 & 0.0437 \\
& 0.05 & 0.0409 & 0.0665 & 0.1006 & 0.0265 & 0.0347 & 0.0432 \\
& 0.09 & 0.0433 & 0.0670 & 0.1010 & 0.0279 & 0.0355 & 0.0441 \\
& 0.10 & 0.0426 & 0.0648 & 0.01004 & 0.0278 & 0.0349 & 0.0439 \\
& 0.15 & 0.0414 & 0.0646 & 0.0990 & 0.0270 & 0.0344 & 0.0431 \\
& Dynamic & \textbf{0.0438} & \textbf{0.0692} & \textbf{0.1050} & \textbf{0.0281} & \textbf{0.0363} & \textbf{0.0453} \\
\midrule
\multirow{5}{*}{Home}
& 0.01 & 0.0118 & 0.0196 & 0.0314 & 0.0075 & 0.0100 & 0.0130 \\
& 0.05 & 0.0117 & 0.0195 & \textbf{0.0319} & 0.0075 & 0.0100 & 0.0131 \\
& 0.10 & 0.0118 & 0.0201 & 0.0315 & 0.0076 & \textbf{0.0103} & 0.0131 \\
& 0.15 & 0.0116 & 0.0193 & 0.0311 & 0.0074 & 0.0099 & 0.0128 \\
& 0.20 & 0.0117 & 0.0195 & 0.0311 & 0.0074 & 0.0099 & 0.0128 \\
& Dynamic & \textbf{0.0124} & \textbf{0.0205} & \textbf{0.0324} & 0.0077 & \textbf{0.0103} & \textbf{0.0133} \\
\bottomrule
\end{tabular}%
}
\end{table}



\noindent\textbf{Impact of Entropy Coefficient.} The entropy regularization term in Equation (13) balances exploration and exploitation during policy learning. We experiment with both fixed and dynamic entropy coefficients, and the results with the SASRec backbone are summarized in Table~\ref{tab:entropy_ablation}. 
The dynamic entropy strategy, which adaptively adjusts the coefficient based on training progress and current policy entropy, consistently achieves the best performance across almost all metrics. This adaptive mechanism maintains higher exploration during early training stages and gradually reduces exploration for refined exploitation in later phases, effectively avoiding premature convergence to suboptimal strategies while preventing excessive randomness. 


\begin{figure}[t]
    \centering
    \includegraphics[width=\columnwidth]{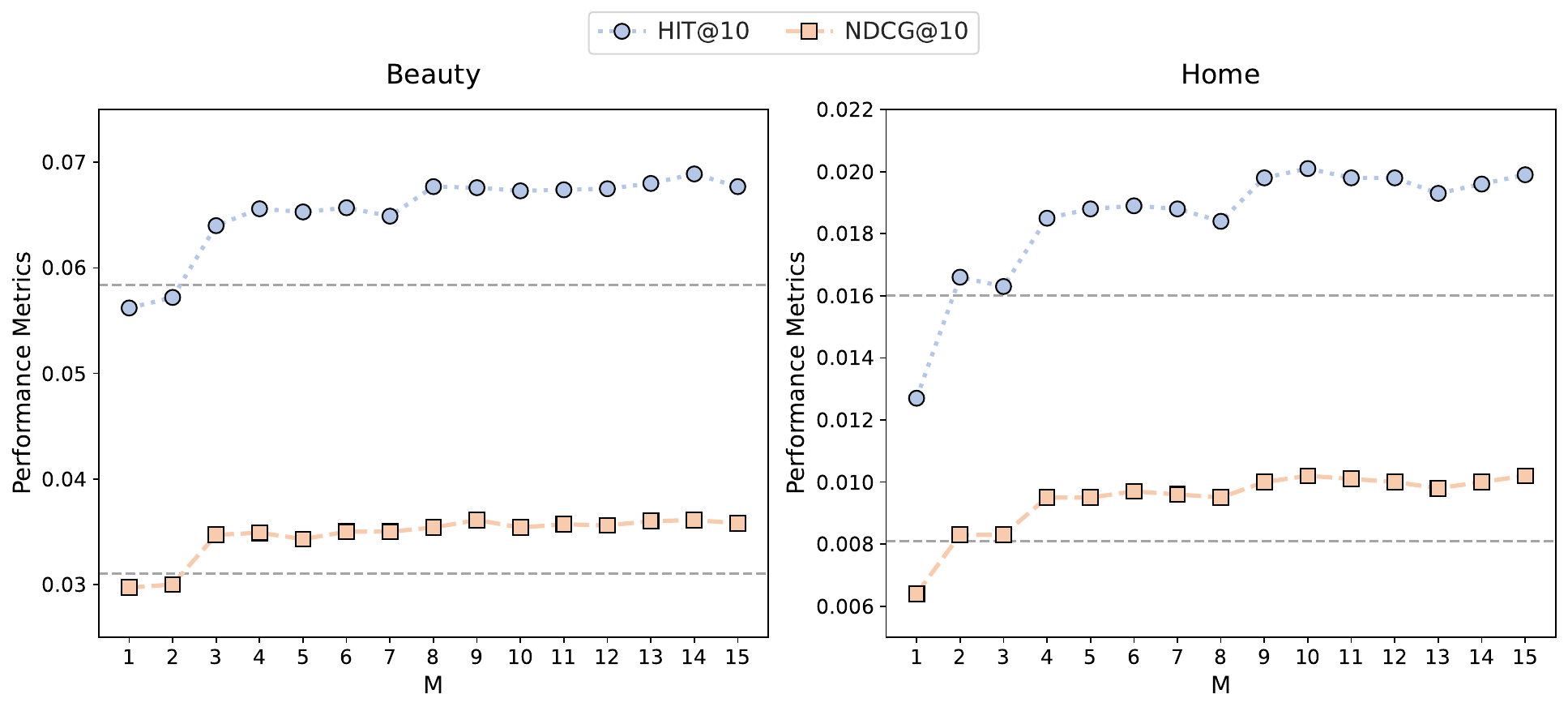}
    \caption{Performance with different augmentation times $m$.}
    \Description{Table showing the effect of augmentation times $m$ on performance metrics HIT@10 and NDCG@10 for Beauty and Home datasets using SASRec model. The base model performance is compared with different $m$ values from 5 to 15.}
    \label{fig:effect_m_compact}
\end{figure}

\noindent\textbf{Impact of Augmentation Times.} We systematically investigate the impact of the number of augmentations $m$ per input sequence. The results with the SASRec backbone are presented in Figure~\ref{fig:effect_m_compact}. Performance on both datasets improves as $m$ increases, but with different patterns. Beauty shows a gradual improvement, reaching its optimum around $m=13$. In contrast, Home exhibits a significant performance jump at $m=10$, after which gains plateau. This indicates that moderately increasing augmentations provides richer predictive signals, but excessive augmentation introduces redundancy. It also reveals dataset-specific sensitivity to augmentation intensity.

\begin{table}[t]
\centering
\caption{Performance with different reward weight configurations $(\alpha, \beta)$.}
\label{tab:reward_weight}
\resizebox{\columnwidth}{!}{%
\begin{tabular}{lcccccccc}
\toprule
\multirow{2}{*}{\textbf{Dataset}} & \multirow{2}{*}{$(\alpha, \beta)$} & \multicolumn{3}{c}{\textbf{HIT}} & \multicolumn{3}{c}{\textbf{NDCG}} \\
\cmidrule(lr){3-5} \cmidrule(lr){6-8}
& & @5 & @10 & @20 & @5 & @10 & @20 \\
\midrule
\multirow{6}{*}{Beauty} 
& (1.0, 0.0) & 0.0414 & 0.0647 & 0.0994 & 0.0264 & 0.0339 & 0.0426 \\
& (0.9, 0.1) & 0.0421 & 0.0674 & 0.1040 & 0.0274 & 0.0356 & 0.0447 \\
& (0.7, 0.3) & 0.0426 & 0.0673 & 0.1035 & 0.0279 & 0.0359 & 0.0449 \\
& (0.6, 0.4) & 0.0436 & 0.0686 & 0.1039 & \textbf{0.0283} & \textbf{0.0363} & 0.0452 \\
& (0.2, 0.8) & \textbf{0.0438} & \textbf{0.0692} & \textbf{0.1050} & 0.0281 & \textbf{0.0363} & \textbf{0.0453} \\
& (0.0, 1.0) & 0.0436 & 0.0664 & 0.1024 & 0.0282 & 0.0355 & 0.0445 \\
\midrule
\multirow{6}{*}{Home}
& (1.0, 0.0) & 0.0109 & 0.0187 & 0.0319 & 0.0071 & 0.0096 & 0.0131 \\
& (0.7, 0.3) & 0.0120 & 0.0200 & 0.0319 & 0.0076 & 0.0102 & 0.0131 \\
& (0.4, 0.6) & 0.0122 & 0.0199 & 0.0315 & \textbf{0.0077} & 0.0102 & 0.0131 \\
& (0.1, 0.9) & \textbf{0.0124} & \textbf{0.0205} & \textbf{0.0324} & \textbf{0.0077} & \textbf{0.0103} & \textbf{0.0133} \\
& (0.5, 0.5) & 0.0118 & 0.0196 & 0.0315 & 0.0075 & 0.0100 & 0.0130 \\
& (0.0, 1.0) & 0.0117 & 0.0197 & 0.0311 & 0.0074 & 0.0100 & 0.0128 \\
\bottomrule
\end{tabular}%
}
\end{table}

\noindent\textbf{Impact of the Reward Weight.} The reward function in Equation (12) combines macro and rank rewards with coefficients $\alpha$ and $\beta$. Table~\ref{tab:reward_weight} presents the results for key configurations with the SASRec backbone. Two observations emerge. First, the best performance is consistently achieved with mixed reward weights rather than using either reward alone, indicating that the two components provide complementary learning signals.
Second, the fine-grained rank reward becomes more important on the sparser Home dataset, where configurations with larger $\beta$ yield superior performance.

\section{CONCLUSION}

In this work, we explore adaptive test-time augmentation for sequential recommendation. We first empirically reveal that a fixed “one-size-fits-all” strategy is suboptimal as it cannot cater to diverse behavioral sequence patterns. 
Based on this analysis, we propose AdaTTA, a reinforcement learning framework that learns to select the optimal augmentation operator for each input sequence adaptively. It formulates the selection as a Markov Decision Process, where a policy network leverages hybrid sequence representations to make real-time decisions, optimized using a joint macro-rank reward design. Extensive experiments shown that our approach achieves consistent improvements over all fixed-strategy baselines in terms of effectiveness, robustness and generalizability, while incurring acceptable computational overhead.



\bibliographystyle{ACM-Reference-Format}
\bibliography{main}  

\end{document}